\documentclass[12pt]{article}
\usepackage{amsmath}
\usepackage{graphicx}
\usepackage{color}
\begin{document}
\baselineskip=18 pt
\begin{center}
{\large{\bf The Dirac Equation in (1+2)-dimensional G\"{u}rses space-time backgrounds}}
\end{center}

\vspace{.5cm}

\begin{center}
{\bf Faizuddin Ahmed}\footnote{faizuddinahmed15@gmail.com ; faiz4U.enter@rediffmail.com}\\ 
{\it Ajmal College of Arts and Science, Dhubri-783324, Assam, India}
\end{center}

\vspace{.5cm}

\begin{abstract}

In this article, we investigate the relativistic quantum dynamics of spin-$\frac{1}{2}$ particles in (1+2)-dimensional G\"{u}rses space-time backgrounds, and analyze the effects on the eigenvalues. We solve the Dirac equation using the Nikiforov-Uvarov method in the considered framework, and evaluate the energy eigenvalues and corresponding wave function.

\end{abstract}

{\it keywords:} rotational symmetry, Relativistic wave equations: Dirac equation, energy spectrum, wave functions, special functions : Nikiforov-Uvarov method.

\vspace{0.1cm}


\section{Introduction}

In modern physics, the Dirac equation is one of the remarkable achievements because of its important role in particle physics by predicting the existence of anti-particles. The Dirac equation is used to describes spin-$\frac{1}{2}$ particles such as fermions and the Klein-Gordon equation for spin-$0$ particles such as mesons. The Dirac equation has been considered in different curved backgrounds due to its considerable applications in astrophysics, cosmology, and condensed matter physics (see, Refs. \cite{SKM,MK,VMV,YS}). To obtain the solutions of the relativistic wave equations, various quantum mechanical techniques such as the Nikiforov-Uvarov \cite{AF} (see, Refs. \cite{CB,YF,HE,HH}), super-symmetric quantum mechanics \cite{FC} (see, Refs. \cite{XZ,XQZ}), asymptotic iteration method (AIM) \cite{HC} etc. have been applied. In this investigations various kinds of potentials including the harmonic oscillator potential \cite{Su}, Eckart potential \cite{CS,XZ}, Woods-Saxon potential \cite{CB}, Hyperbolic-type potential \cite{AA}, scalar or vector potential (linear and/or of Coulomb-type) have been considered.

The relativistic quantum effects on spin-$0$ and spin-half particles in G\"{o}del as well as G\"{o}del-type geometries have been addressed by several authors. In Ref. \cite{BDF}, the authors studied the Klein-Gordon and Dirac equations in a class of G\"{o}del-type geometries with zero, positive and negative curvatures. In Ref. \cite{ND}, the relativistic quantum dynamics of spin-$0$ particles in G\"{o}del-type geometries, were investigated. They solved the equation and analyzed the similarity of the energy eigenvalues with the Landau levels in flat, spherical and hyperbolic spaces (see also, Ref. \cite{ND2}). In Ref. \cite{SD}, the Klein-Gordon equation is solved in the background of a class of flat G\"{o}del-type geometries and analyzed the similarity of energy eigenvalues with the Landau levels \cite{LL}. In Ref. \cite{LCNS}, the Klein-Gordon equation with vector and scalar potentials of Coulomb-type under the influence of non-inertial effects in the cosmic string space-time, were investigated. There it has been shown that the presence of potential parameter and cosmic string allows the formation of bound states solutions and the energy spectrum get modifies. In Ref. \cite{LCNS2}, two different classes of solutions for the Klein-Gordon equation in the presence of the Coulomb scalar potential under the influence of non-inertial effects in the cosmic string space-time, were studied. It has been shown that the presence of the potential parameter allows the formation of bound states and the energy spectrum of the scalar particles depends on the cosmic string parameter. In Ref. \cite{EPJC}, the relativistic quantum effect on scalar particles in a class of topologically trivial flat G\"{o}del-type geometries, was investigated. We have shown that the energy spectrum gets modified and depends on the vorticity parameter characterising the space-time. In Rf. \cite{EPJC2}, the linear confinement of scalar particles in a class of topologically trivial flat G\"{o}del-type geometries, was investigated. We have obtained a compact expression of the energy spectrum which depends on different parameters. Also spin-$0$ massive charged particles in the presence of vector and scalar potentials of Coulomb-type, were studied there. In Ref. \cite{EPJC3}, the Dirac particles in a class of topologically trivial flat G\"{o}del-type geometries, was investigated. In Ref. \cite{JC}, the relativistic quantum dynamics of spin-$0$ particles in a class of G\"{o}del-type geometries with a cosmic string, were studied. They solved the relativistic wave equation and analyzed the similarity of the energy eigenvalues with the Landau levels in flat, spherical and hyperbolic spaces. It was shown that the presence of the cosmic string, and the vorticity parameter modifies the energy levels and breaks their degeneracy. In Ref. \cite{JC2}, the quantum influence of topological defects in G\"{o}del-type geometries in the flat, spherical and hyperbolic cases, was investigated. In Ref. \cite{GQG}, the relativistic quantum dynamics of Dirac particle with topological defect in a class of G\"{o}del-type geometries with torsion was investigated. In Ref. \cite{GQG2}, Weyl fermions in G\"{o}del-type geometries with a topological defect, were investigated. In Ref. \cite{SGF}, scalar quantum particles in the backgrounds of Kerr-Newman, G\"{o}del and Friedmann-Robertson-Walker (FRW) space-times with a cosmic string, were investigated. In Ref. \cite{AH}, the photon equation (mass-less Duffin-Kemmer-Petiau equation) has been written explicitly for the general type of stationary G\"{o}del and G\"{o}del-type geometries and solved exactly. There, the harmonic oscillator behaviour of the solutions was discussed and the energy spectrum of the photon was obtained. In Ref. \cite{LCNS3}, the relativistic wave equation for spin-$\frac{1}{2}$ particles in the background of Melvin space-time, a space-time where the metric is determined by a magnetic field, were investigated. In Ref. \cite{MH}, the Dirac oscillator in the background of a spinning cosmic string space-time, were investigated. In Ref. \cite{MM} the relativistic dynamics of a Dirac field in the Som-Raychaudhuri space-time, and a class of flat G\"{o}del-type geometries, were investigated. In Ref. \cite{EPJC4}, Dirac fermions in the Som–Raychaudhuri space-time with scalar and vector potential and the energy momentum distributions, were investigated. All the above works have been done in four-dimensional space-times. In addition, the relativistic wave equations are also investigated in two- and three-dimensional space-time. In Ref. \cite{HF}, the Dirac equation in two-,  three-dimensional spaces was studied in the presence of a magnetic field. In two-dimensional space, an imaginary value for the radial component of the electromagnetic potential was considered and different angular gauge fields obtained in terms of the master function. The bound state solutions of the Dirac equation for the one-dimensional linear potential with Lorentz scalar and vector couplings, were investigated in Ref. \cite{ASDC}. The solutions of the Dirac equation and the non-relativistic limit for a linear scalar potential, were studied in Ref. \cite{JRH}. The three-dimensional Dirac oscillator in a thermal bath was studied in Ref. \cite{MHP}. In Ref. \cite{AOP}, the relativistic quantum dynamics of spin-$0$ particles in (1+2)-dimensional G\"{u}rses space-time was studied. The Klein-Gordon equation without interaction is solved in the considered framework and the energy spectrum and corresponding wave functions obtained in detail. There, the fact that the presence of the parameter $\Omega$ modifies the energy eigenvalues was shown and thta, for massless scalar particles, the energy eigenvalues are similar to the Landau levels. In Ref. \cite{AOP2}, the Klein-Gordon oscillator is solved in the backgrounds of (1+2)-dimensional G\"{u}rses space-time using the Nikiforov-Uvarov (NU) method. In Ref. \cite{GERG}, the generalised Klein-Gordon oscillator subject to Coulomb-type scalar potential in the backgrounds of (1+2)-dimensional G\"{u}rses space-time was investigated. In Ref. \cite{MDO}, exact solutions of the Dirac equation on a static curved space-time with electromagnetic interactions was studied. In Ref. \cite{JC3}, the Dirac equation for the Dirac oscillator in the background space-time of a cosmic string, of a magnetic cosmic string, and of a cosmic dislocation was studied.

In this work, we study the solution for Dirac equation in the backgrounds of (1+2)-dimensional rotational symmetry space-time with rotation. We solve the Dirac equation using the Nikiforov-Uvarov method in curved space backgrounds, and obtain the energy eigenvalues and corresponding wave functions.

\section{Three-dimensional G\"{u}rses space-time}

Consider the following stationary and rotational symmetry (1+2)-dimensional G\"{u}rses metric \cite{Gur} (see refs. \cite{AOP,AOP2,GERG} and appendix A)
\begin{eqnarray}
ds^2&=&-dt^2+dr^2+r^2\,(1-\Omega^2\,r^2)\,d\phi^2-2\,\Omega\,r^2\,dt\,d\phi,\nonumber\\
&=&g_{\mu\nu}\,dx^{\mu}\,dx^{\nu},
\label{1}
\end{eqnarray}
where $\mu,\nu=0,1,2$. The metric (\ref{1}) can be express as
\begin{equation}
ds^2=-(dt+H\,d\phi)^2+D^2\,d\phi^2+dr^2,
\label{2}
\end{equation}
where $H (r)=\Omega\,r^2$ and $D (r)= r$. The properties of the above space-time has been studied in detail in Ref. \cite{AOP}. In this work, we investigate the relativistic quantum motion of Dirac fields, and analyze the effects on the energy eigenvalues.

The above space-time satisfies the following conditions (see Eq. (12) in Ref. \cite{EPJC2} and related references therein):
\begin{equation}
    \frac{H'}{D}=2\,\Omega\quad,\quad \frac{D''}{D}=0,
    \label{3}
\end{equation}
which clearly indicates that the metric belongs to flat class of G\"{o}del-type geometries (which is also called zero curvature) in the three-dimensional case, where the prime denotes derivative w. r. t. $r$. The properties of three dimensional metrics have been studied in detail in refs. \cite{RJG,MG,MG2}.

The metric has signature $(-,+,+)$ and the determinant of the corresponding metric tensor $g_{\mu\nu}$ is
\begin{equation}
det\;g=-r^2.
\label{4}
\end{equation}
The coordinates fulfil the following ranges
\begin{equation}
-\infty < t < \infty,\quad 0\leq r,\quad 0\leq \phi \leq 2\pi.
\label{5}
\end{equation}
The covariant form of the metric tensor for the space-time (\ref{1}) is
\begin{equation}
g_{\mu\nu} ({\bf x})=\left (\begin{array}{lll}
-1 & 0 & \quad-\Omega\,r^2 \\
\quad 0 & 1 & \quad\quad 0 \\
-\Omega\,r^2 & 0 & r^2\,(1-\Omega^2\,r^2)
\end{array} \right)
\label{6}
\end{equation}
with its inverse 
\begin{equation}
g^{\mu\nu} ({\bf x})=\left (\begin{array}{lll}
-1+\Omega^2\,r^2 & 0 & -\Omega \\
\quad\quad 0 & 1 & \quad 0 \\
\quad-\Omega & 0 & \quad \frac{1}{r^2} 
\end{array} \right).
\label{7}
\end{equation}
The Christoffel symbols of the metric (\ref{1}) are
\begin{eqnarray}
\Gamma^{0}_{01}&=&\Omega^2\,r\quad,\quad \Gamma^{0}_{12}=\Omega^3\,r^3\quad,\quad \Gamma^{1}_{20}=\Omega\,r,\nonumber\\
\Gamma^{1}_{22}&=&r\,(-1+2\,\Omega^2\,r^2)\quad,\quad \Gamma^{2}_{01}=-\frac{\Omega}{r}\quad,\quad \Gamma^{2}_{12}=\frac{1}{r}-\Omega^2\,r.
\label{8}
\end{eqnarray}

Using the definition of $e^{\mu}_{(a)}$ and $e^{(a)}_{\mu}$ in space-time (\ref{1}), we have
\begin{equation}
e^{(a)}_{\mu} ({\bf x})=\left (\begin{array}{lll}
1 & 0 & \Omega\,r^2 \\
0 & 1 & \quad 0 \\
0 & 0 & \quad r 
\end{array} \right)\quad,\quad
e^{\mu}_{(a)} ({\bf x})=\left (\begin{array}{lll}
1 & 0 & -\Omega\,r \\
0 & 1 & \quad 0 \\
0 & 0 & \quad \frac{1}{r} 
\end{array} \right)
\label{9}
\end{equation}
which must satisfy
\begin{eqnarray}
&&e^{\mu}_{(a)} ({\bf x})\,e^{(a)}_{\nu} ({\bf x})=\delta^{\mu}_{\nu},\nonumber\\
&&e^{(a)}_{\mu} ({\bf x})\,e^{\mu}_{(b)} ({\bf x})=\delta^{a}_{b},\nonumber\\
&&g_{\mu\nu} ({\bf x})= e^{(a)}_{\mu} ({\bf x})\,e^{(b)}_{\nu} ({\bf x})\,\eta_{(a)(b)},
\label{10}
\end{eqnarray}
where $\eta_{(a)(b)}=\mbox{diag} (-1, 1, 1)$ and $a,b=0,1,2$.

\subsection{\bf The Dirac Equation in the backgrounds of G\"{u}rses metric}

The Dirac equation for a free Fermi field of mass $m$ in curved space-time can be written as
\begin{equation}
[i\,\gamma^{\mu} ({\bf x}) \nabla_{\mu}-m]\Psi=0,
\label{11}
\end{equation}
where $\nabla_{\mu}$ is given by
\begin{equation}
\nabla_{\mu}=\partial_{\mu}+\Gamma_{\mu}.
\label{12}
\end{equation}
Here $\Gamma_{\mu}$ is the spinorial affine connection
\begin{equation}
\Gamma_{\mu}=-\frac{1}{8}\,\omega_{\mu\,(a)(b)}\,[\gamma^{a},\gamma^{b}].
\label{13}
\end{equation}

The Dirac matrices $\gamma^{a}$ related with the Pauli matrices are
\begin{equation}
\gamma^{0}=\left (\begin{array}{ll} 
{\bf \sigma^0} & \,\,{\bf 0} \\
{\bf 0} & -{\bf \sigma^0}
\end{array} \right),\quad
\gamma^{1}=\left (\begin{array}{ll}
\quad {\bf 0} & \sigma^{1} \\
-\sigma^{1} & {\bf 0} 
\end{array} \right),\nonumber
\end{equation}
\begin{equation}
\gamma^{2}=\left (\begin{array}{ll}
\quad {\bf 0} & \sigma^{2} \\
-\sigma^{2} & {\bf 0} 
\end{array} \right),\quad
\gamma^{3}=\left (\begin{array}{ll}
\quad {\bf 0} & \sigma^{3} \\
-\sigma^{3} & {\bf 0} 
\end{array} \right),
\label{14}
\end{equation}
where ${\bf 0}$ is $2\times 2$ null matrix and the Pauli matrices are 
\begin{equation}
\sigma^{0}=\left (\begin{array}{ll} 
1 & 0 \\
0 & 1
\end{array} \right),\quad
\sigma^{1}=\left (\begin{array}{ll} 
0 & 1 \\
1 & 0
\end{array} \right),\quad
\sigma^{2}=\left (\begin{array}{ll}
0 & -i \\
i & \,0
\end{array} \right),\quad 
\sigma^{3}=\left (\begin{array}{ll}
1 & \,0 \\
0 & -1
\end{array} \right).
\label{15}
\end{equation}

The generalised Dirac matrices $\gamma^{\mu} ({\bf x})=e^{\mu}_{a} ({\bf x})\,\gamma^{a}$ are
\begin{equation}
\gamma^{t} ({\bf x})=\left(\begin{array}{ll} 
{\bf I} & -\Omega\,r\,\sigma^2 \\
\Omega\,r\,\sigma^2 & -{\bf I}
\end{array} \right),\nonumber\\
\end{equation}
\begin{equation}
\gamma^{r} ({\bf x})=\left (\begin{array}{ll}
\quad{\bf 0} & \sigma^1 \\
-\sigma^1 & {\bf 0} 
\end{array} \right),\nonumber\\
\end{equation}
\begin{equation}
\gamma^{\phi} ({\bf x})=\frac{1}{r}\,\left (\begin{array}{ll}
\quad{\bf 0} & \sigma^2 \\
-\sigma^2 & {\bf 0} 
\end{array} \right).
\label{16}
\end{equation}

The spin connections $\omega_{\mu\,(a)(b)}$ are defined by 
\begin{equation}
\omega_{\mu\,{(a)}{(b)}} ({\bf x})=\eta_{(a)(c)}\,e^{(c)}_{\nu}\,e^{\tau}_{(b)}\,\Gamma^{\nu}_{\tau\mu}-\eta_{(a)(c)}\,e^{\nu}_{(b)}\,\partial_{\mu}\,e^{(c)}_{\nu}.
\label{17}
\end{equation}
And these are
\begin{equation}
\omega_{t\,{(a)}{(b)}} ({\bf x})=\Omega\,\left (\begin{array}{lll}
0 & \quad 0 & 0 \\
0 & \quad 0 & 1  \\
0 & -1 & 0 
\end{array} \right),\nonumber\\
\end{equation}
\begin{equation}
\omega_{r\,{(a)}{(b)}} ({\bf x})=\Omega\left (\begin{array}{lll}
\quad 0 & 0 & 1 \\
\quad 0 & 0 & 0 \\
-1 & 0 & 0
\end{array} \right),\\
\label{18}
\end{equation}
\begin{equation}
\omega_{\phi\,{(a)}{(b)}} ({\bf x})=\left (\begin{array}{lll}
0 & \,\,-\Omega\,r & \quad\quad 0 \\
\Omega\,r & \quad 0 & -1+\Omega^2\,r^2 \\
0 & 1-\Omega^2\,r^2 & \quad\quad 0
\end{array} \right).\nonumber
\end{equation}

The components of spinorial affine connection $\Gamma_{\mu} (\bf x)$ are
\begin{equation}
\Gamma_{t} ({\bf x})=\frac{i\,\Omega}{2}\,\left (\begin{array}{ll}
\sigma^3 & {\bf 0} \\
{\bf 0} & \sigma^3
\end{array} \right),\nonumber
\end{equation}
\begin{equation}
\Gamma_{r} ({\bf x})=-\frac{\Omega}{2}\,\left (\begin{array}{ll}
{\bf 0} & \sigma^2 \\
\sigma^2 & {\bf 0}
\end{array} \right),\nonumber
\end{equation}
\begin{equation}
\Gamma_{\phi} ({\bf x})=\frac{\Omega\,r}{2}\,\left (\begin{array}{ll}
{\bf 0} & \sigma^1 \\
\sigma^1 & {\bf 0}
\end{array} \right)-\frac{i\,(1-\Omega^2\,r^2)}{2}\,\left (\begin{array}{ll}
\sigma^3 & {\bf 0} \\
{\bf 0} & \sigma^3
\end{array} \right),
\label{19}
\end{equation}

The second term in Eq. (\ref{11}) using (\ref{12}) becomes
\begin{equation}
\gamma^{\mu} ({\bf x})\,\Gamma_{\mu} ({\bf x})=\frac{1}{2}\,\left (\begin{array}{ll}
i\,\Omega\,\sigma^3 & \frac{\sigma^1}{r} \\
-\frac{\sigma^1}{r} & -i\,\Omega\,\sigma^3
\end{array} \right).
\label{20}
\end{equation}
Therefore from Eq. (\ref{11}), we have
\begin{equation}
[i\,(\gamma^{t}\,\partial_{t}+\gamma^{r}\,\partial_{r}+\gamma^{\phi}\,\partial_{\phi})+i\,\gamma^{\mu}\,\Gamma_{\mu}-m]\,\Psi=0.
\label{21}
\end{equation}
We write the Dirac spinor as
\begin{equation}
\Psi (t,r,\phi)=e^{-i\,E\,t}\,e^{j\,\phi}\,\left (\begin{array}{c}
\psi_1 \\
\psi_2
\end{array} \right),
\label{22}
\end{equation}
where $E=-i\,\partial_{t}$ is the total energy, and $j=l+\frac{1}{2}$ with $l=0,\pm 1,\pm 2...$ is the orbital angular momentum operator, and $\sqrt{-1}=i$.

Substituting the ansatz (\ref{22}) into Eq. (\ref{21}) and using Eqs. (\ref{15}), (\ref{20}), we get
\begin{eqnarray}
E\left (\begin{array}{c}
\psi_1-\Omega\,r\,\sigma^2\,\psi_2 \\
-\psi_2+\Omega\,r\,\sigma^2\,\psi_1 
\end{array} \right) &+&\left (\begin{array}{c}
i\,\sigma^1\,\psi'_2 \\
-i\,\sigma^1\,\psi'_1
\end{array} \right )-\frac{j}{r}\,\left (\begin{array}{c}
\sigma^2\,\psi_2 \\
-\sigma^2\,\psi_1
\end{array} \right)\nonumber\\
&+&\frac{i}{2} \left (\begin{array}{c}
i\,\Omega\,\sigma^3\,\psi_1+\frac{\sigma^1}{r}\,\psi_2 \\
-i\,\Omega\,\sigma^3\,\psi_2-\frac{\sigma^1}{r}\,\psi_1
\end{array} \right)=m\left (\begin{array}{c}
\psi_1 \\
\psi_2
\end{array} \right).
\label{23}
\end{eqnarray}
Therefore we have the following differential equations:
\begin{eqnarray}
\label{24}
&&(E-\frac{\Omega}{2}\,\sigma^3-m)\,\psi_1=[(\Omega\,E\,r+\frac{j}{r})\,\sigma^2-\frac{i}{2\,r}\,\sigma^1]\,\psi_2-i\,\sigma^1\,\psi'_{2},\\
\label{25}
&&(E-\frac{\Omega}{2}\,\sigma^3+m)\,\psi_2=[(\Omega\,E\,r+\frac{j}{r})\,\sigma^2-\frac{i}{2\,r}\,\sigma^1]\,\psi_1-i\,\sigma^1\,\psi'_{1},
\end{eqnarray}
After decoupling, we get the following second-order differential equation:
\begin{equation}
\psi''_{s}+\frac{1}{r}\,\psi'_{s}+[\lambda-\Omega^2\,E^2\,r^2-\frac{\tilde{j}^2}{r^2}]\,\psi_{s}=0,
\label{26}
\end{equation}
where
\begin{eqnarray}
\lambda&=&E^2+\frac{\Omega^2}{4}-m^2-2\,\Omega\,E\,(j+s),\nonumber\\
\tilde{j}^2&=&j^2-j\,s+\frac{1}{4}=(j-\frac{s}{2})^2.
\label{27}
\end{eqnarray}
Note that $\psi_1,\psi_2$ are wave functions of $\sigma^3$ with eigenvalues $\pm 1$, so we can write $\psi_{s}=(\psi_{+},\psi_{-})^T$ with $\sigma^3\,\psi_{s}=s\,\psi_{s}$, $s=\pm 1$. Here we have used
\begin{eqnarray}
(E-\frac{\Omega}{2}\,\sigma^3)^2\,\psi_{s}&=&E^2\,\psi_{s}+\frac{\Omega^2}{4}\,\psi_{s}-\Omega\,E\,\sigma^3\,\psi_{s}\nonumber\\
&=&(E^2+\frac{\Omega^2}{4}-\Omega\,E\,s)\,\psi_{s}.
\label{28}
\end{eqnarray}
We employ the change of variable $x=\Omega\,E\,r^2$, then rewrite the radial Eq. (\ref{26}) in the form
\begin{equation}
\psi''_{s} (x)+\frac{1}{x}\,\psi'_{s} (x)+\frac{1}{x^2}\,(-\xi_1\,x^2+\xi_2\,x-\xi_3)\,\psi_{s} (x)=0,
\label{29}
\end{equation}
where
\begin{equation}
\xi_1=\frac{1}{4}\quad,\quad \xi_2=\frac{\lambda}{4\,\Omega\,E}\quad ,\quad \xi_3=\frac{\tilde{j}^2}{4}.
\label{30}
\end{equation}
The Eq.(\ref{29}) is the second-order Nikiforov-Uvarov differential equation.

Therefore, the energy eigenvalues equation \cite{AOP2} is given by
\begin{eqnarray}
&&(2\,n+1)\,\sqrt{\xi_1}-\xi_2+2\,\sqrt{\xi_1\,\xi_3}=0\nonumber\\ \Rightarrow
&&\lambda=2\,(2\,n+1+|\tilde{j}|)\,\Omega\,E\nonumber\\\Rightarrow
&&E^2_{n,l}-2\,\Omega\,E_{n,l}\,(2\,n+1+|\tilde{j}|+j+s)-m^2+\frac{\Omega^2}{4}=0
\label{31}
\end{eqnarray}
with the energy spectrum
\begin{eqnarray}
\label{32}
E_{n,l}&=&\Omega\,(2\,n+1+|\tilde{j}|+j+s)\pm \sqrt{\Omega^2\,(2\,n+1+|\tilde{j}|+j+s)^2+m^2-\frac{\Omega^2}{4}}\nonumber\\
&=&\Omega\,(2\,n+1+|j-\frac{s}{2}|+j+s)\pm \sqrt{\Omega^2\,(2\,n+1+|j-\frac{s}{2}|+j+s)^2+m^2-\frac{\Omega^2}{4}}\nonumber\\
&=&\Omega\,\{2\,n+1+|l+\frac{1}{2}(1-s)|+l+\frac{1}{2}+s \}\nonumber\\
&&\pm \sqrt{\Omega^2\,(2\,n+1+|l+\frac{1}{2}(1-s)|+l+\frac{1}{2}+s)^2+m^2-\frac{\Omega^2}{4}}.
\end{eqnarray}
where $n=0,1,2,3.....$, $l=0,\pm 1,\pm 2...$ and $s=\pm 1$. 

The corresponding wave functions \cite{AOP2} is
\begin{equation}
\psi_{s\,n,l} (x)=|N|\,x^{\frac{1+|l+\frac{1}{2}(1-s)|}{4}}\,e^{-\frac{x}{2}}\,L^{(|l+\frac{1}{2}(1-s)|)}_{n} (x),
\label{33}
\end{equation}
where $|N|$ is the normalization constant and $L^{(a)}_{n} (x)$ is the generalized Laguerre polynomials.

For massless Dirac fermions, that is, $m=0$, we have the following energy eigenvalues:
\begin{eqnarray}
E_{n,l}&=&\Omega\,[(2\,n+1+|l+\frac{1}{2}(1-s)|+l+\frac{1}{2}+s)\nonumber\\
&&+ \sqrt{(2\,n+1+|l+\frac{1}{2}(1-s)|+l+\frac{1}{2}+s)^2-\frac{1}{4}}].
\label{34}
\end{eqnarray}
We have seen that for massless Dirac fermions in the backgrounds of G\"{u}rses space-time, the presence of the term $\frac{\Omega^2}{4}$ causes an asymmetric in the energy eigenvalues and shifts the energy levels.

The energy eigenvalues of massless Dirac fermions in (1+3)-dimensional space-time in ref. \cite{GQG} (see ref. \cite{BDF}) with torsion field ($S^z$) and topological defect are ($k=0=M$, see Eq. (39) in Ref. \cite{GQG})
\begin{eqnarray}
    E_{n,j}&=&2\,\Omega\,(n+\frac{1}{2}+\frac{|j|}{2\,\alpha}+\frac{j}{2\,\alpha})\nonumber\\
    &&\pm \sqrt{4\,\Omega^2\,(n+\frac{1}{2}+\frac{|j|}{2\,\alpha}+\frac{j}{2\,\alpha})^2+(\frac{S^z}{8})^2}.
    \label{35}
\end{eqnarray}
where $j=l+\frac{1}{2}$. For the torsion free field and without topological defects ($\alpha=1$), the energy eigenvalues reduce to
\begin{equation}
    E_{n,l}=2\,\Omega\,(2\,n+1+|l+\frac{1}{2}|+l+\frac{1}{2}).
    \label{36}
\end{equation}

Furthermore, the eigenvalues of the energy of Weyl fermions in the four-dimensional Som-Raychaudhuri metric with a topological defect in Ref. \cite{GQG2} (see Eq. (26) there) are given by
\begin{eqnarray}
    E_{n,j}&=&2\,\Omega\,(n+\frac{1}{2}+\frac{|j|+j}{2\,\alpha})\nonumber\\
    &&\pm \sqrt{4\,\Omega^2\,(n+\frac{1}{2}+\frac{|j|+j}{2\,\alpha})^2+k^2}.
    \label{37}
\end{eqnarray}
where $j=l+\frac{1}{2}$ is a half-integer. If one chooses $k=0$ and $\alpha=1$, then the energy of a particle confined to plane without topological defects from Eq. (\ref{37}) is given by
\begin{equation}
    E_{n,l}=2\,\Omega\,(2\,n+1+|l+\frac{1}{2}|+l+\frac{1}{2}).
    \label{38}
\end{equation}

In Ref. \cite{Hassan}, the relativistic quantum dynamics of the Dirac field in a class of flat G\"{o}del-type space-time (Som-Raychaudhuri metric) with topological defects, are studied. The energy eigenvalues (setting $k=0$ there) are given by (see Eqs. (3.18)--(3.19) in Ref. \cite{Hassan})
\begin{eqnarray}
E_{n,l}&=&2\,\Omega\,(n+\frac{1}{2}+\frac{l}{\alpha})\nonumber\\
&&+\sqrt{4\,\Omega^2\,(n+\frac{1}{2}+\frac{l}{\alpha})^2+m^2+\frac{\Omega^2}{4}+m\,\Omega},\quad E_{n,l}>0.
\label{39}
\end{eqnarray}
\begin{eqnarray}
E_{n,l}&=&2\,\Omega\,(-n-\frac{1}{2}+\frac{l}{\alpha})\nonumber\\
&&+\sqrt{4\,\Omega^2\,(n+\frac{1}{2}-\frac{l}{\alpha})^2+m^2+\frac{\Omega^2}{4}+m\,\Omega},\quad E_{n,l}<0.
\label{40}
\end{eqnarray}
For a massless Dirac field without topological defects ($\alpha=1$), the positive energy eigenvalues from Eq. (\ref{39}) become
\begin{equation}
    E_{n,l}=2\,\Omega\,[(n+\frac{1}{2}+l)+\sqrt{(n+\frac{1}{2}+l)^2+\frac{1}{16}}].
    \label{41}
\end{equation}

In our case, substituting $s=1$ into the Eq. (\ref{34}), we get the energy eigenvalues
\begin{equation}
    E_{n,l}=\Omega\,[(2\,n+|l|+l+\frac{5}{2})+\sqrt{(2\,n+|l|+l+\frac{5}{2})^2-\frac{1}{4}}].
    \label{42}
\end{equation}
Similarly, for $s=-1$ from Eq. (\ref{34}), we get the energy eigenvalues
\begin{equation}
    E_{n,l}=\Omega\,[(2\,n+|l+1|+l+\frac{1}{2})+\sqrt{(2\,n+|l+1|+l+\frac{1}{2})^2-\frac{1}{4}}].
    \label{43}
\end{equation}

From the above analysis it is clear that our result are different from the results obtained in Ref. \cite{GQG,GQG2,Hassan}. Also we have seen that the presence of the term $\frac{\Omega^2}{4}$ causes an asymmetry in the energy eigenvalues and shifts the energy levels which are no longer of Landau type \cite{LL}.

\section{Conclusions}

In Ref. \cite{GQG}, the authors solved the Weyl equation for a family of G\"{o}del-type geometries with a topological defect. They obtained the corresponding Weyl equations in these family of geometries, and solved them exactly. In Ref. \cite{GQG2}, the authors investigated the behaviour of a Dirac particle in the same family of space-times with a topological defect in the Einstein-Cartan theory. They obtained the corresponding Dirac equations in these family of geometries with torsion that contain a cosmic string that passes through the z-axis, and then solved them analytically. In both refs. \cite{GQG,GQG2}, the Som-Raychaudhury metric with a topological defect was considered and the allowed energies for these relativistic quantum systems was obtained and an analogy between these relativistic energy levels and the Landau levels was shown. It was also shown that the presence of topological defect breaks the degeneracy of these relativistic energy levels. In Ref. \cite{GQG}, It was shown that there exists a contribution of the torsion term $\frac{S^z}{8}$ in the allowed energies (see Eq. (40) there). The torsion effect on the allowed energies corresponds to the splitting of each energy level in a doublet. Besides, the torsion contribution is additive due to the constant of motion associated with the operator $\hat {C}$ (see Eq. (22) in Ref. \cite{GQG}). They finally observed the influence of the curvature, the torsion and the topology of the defect of a family of G\"{o}del-type space-times pierced by a topological defect in the energy eigenvalues.

In this work, we have solved the Dirac equation in curved background space in a three-dimensional rotational symmetry space-time. Here we have constructed the Dirac matrices ($\gamma^{\mu} ({\bf x})$) in the considered framework. Consequently, we have constructed the Christoffel symbol, the spin connection and the spinorial affine connection. Having these elements, we could derive the Dirac equation in (1+2)-dimensional G\"{u}rses space-time backgrounds. In {\it sub-section 2.1}, we have solved the Dirac equation in curved space in the considered framework using the Nikiforov-Uvarov method, and evaluated the energy eigenvalues expression Eq. (\ref{32}) and the corresponding wave functions Eq. (\ref{33}). Here we have used the Dirac spinor method as $\sigma^3\,\psi_{+}=\psi_{+}$ and $\sigma^3\,\psi_{-}=-\psi_{-}$ so that $\sigma^3\,\psi_{s}=s\,\psi_{s}$, $s=\pm 1$ where, $\psi_{s}=(\psi_{+},\psi_{-})^T$. For massless Dirac fermions, the energy spectrum reduces to Eq. (\ref{34}) which is different from the result obtained in Refs. \cite{GQG,BDF,GQG2,Hassan} for four-dimensional curved space-time backgrounds. We have seen in Eq. (\ref{34}) that the presence of the term $\frac{\Omega^2}{4}$ causes an asymmetry in the energy eigenvalues and the energy level are shifted. Therefore the obtained energy levels are no longer of Landau type. Note that in our solution the only effect is the vorticity parameter, $\Omega$, without topological defect. We claim that the tool employed here to investigate the quantum dynamics of massless Dirac fermions in three-dimensional G\"{u}rses space-time without torsion, and topological defects are more suitable over the four-dimensional space-time.

\section*{Acknowledgement} The author would like to thank the anonymous kind referees for their valuable comments and suggestions which have greatly improved the present work.

\appendix

\section*{Appendix A : (1+2)-dimensional G\"{u}rses metric}

\setcounter{equation}{0}
\renewcommand{\theequation}{A.\arabic{equation}}

The (1+2)-dimensional space-time considered by G\"{u}rses in \cite{Gur} (see Eq. (5), notations are the same) is given by
\begin{equation}
ds^2=-\phi\,dt^2+2\,q\,dt\,d\theta+\frac{-q^2+h^2\,\psi}{\phi}\,d\theta^2+\frac{1}{\psi}\,dr^2,
\label{A.1}
\end{equation}
where
\begin{eqnarray}
&&\phi=a_0\quad,\quad \psi=b_0+\frac{b_1}{r^2}+\frac{3\,\lambda_0}{4}\,r^2\quad,\quad h=e_0\,r,\nonumber\\
&&q=c_0+\frac{e_0\,\mu}{3}\,r^2\quad,\quad \lambda_0=\lambda+\frac{\mu^2}{27}
\label{A.2}
\end{eqnarray}
and where $a_0,b_0,b_1,c_0,e_0$ are arbitrary constants.

Setting the constants
\begin{equation}
b_1=0\quad,\quad c_0=0\quad,\quad \lambda_0=0
\label{A.3}
\end{equation}
into the metric (\ref{A.1}), we arrive the following metric
\begin{equation}
ds^2=-a_0\,dt^2+\frac{2\,e_0\,\mu}{3}\,r^2\,dt\,d\theta+\frac{1}{a_0}\,(e^{2}_{0}\,b_0\,r^2-\frac{e^{2}_{0}\,\mu^{2}}{9}\,r^4)\,d\theta^2+\frac{1}{b_0}\,dr^2.
\label{A.4}
\end{equation}
Finally choosing the constants
\begin{equation}
a_0=1\quad,\quad b_0=1\quad,\quad e_0=1\quad,\quad \Omega=-\frac{\mu}{3}
\label{A.5}
\end{equation}
into the metric (\ref{A.4}), we get the studied space-time (\ref{1}) presented in this work.


\begin{thebibliography}{99}

\bibitem{SKM} S. K. Moayedi and F. Darabi, Phys. Lett. {\bf A 322}, 173 (2004).
\bibitem{MK} M. Khorrami, M. Alimohammadi, and A. Shariati, Ann. Phys. (N. Y.) {\bf 304}, 91 (2003).
\bibitem{VMV} V. M. Villalba, Mod. Phys. Lett. A {\bf 8}, 2351 (1993).
\bibitem{YS} Y. Sucu and N. Unal, Jour. Math. Phys. {\bf 48}, 052503 (2007).
\bibitem{AF} A. F. Nikiforov and V. B. Uvarov, {\it Special functions of mathematical physics}, Birkhauser, Basel (1988).
\bibitem{CB} C. Berkdemir, A. Berkdemir, and R. Sever, J. Phys. A : Math. Gen. {\bf 39}, 13455 (2006).
\bibitem{YF} Y. F. Cheng and T. Q. Dai, Chinese J. Phys. {\bf 45}, 480 (2007).
\bibitem{HE} H. Egrifes and R. Sever, Phys. Lett A {\bf 344}, 117 (2005).
\bibitem{HH} H. Hassanabadi, E. Maghsoodi, A. N. Ikot, and S. Zarrinkamar, App. Math. Computation {\bf 219}, 9388 (2013).
\bibitem{FC} F. Cooper, A. Khare, U. Sukhatme, Phys. Rep. {\bf 251}, 267 (1995).
\bibitem{XZ} X. Zou, L. -Z. Yi, and C. -S. Jia, Phys. Lett. A {\bf 346}, 54 (2005).
\bibitem{XQZ} X. -Q. Zhao, C. -S. Jia, and Q. -B. Yang, Phys. Lett. A {\bf 337}, 189 (2005).
\bibitem{HC} H. Ciftci, R. L. Hall, N. Saad, J. Phys. A : Math. Gen. {\bf 38}, 1147 (2005).
\bibitem{Su} R. K. Su and Z. Q. Ma, J. Phys. A : Math. Gen. {\bf 19}, 1739 (1986).
\bibitem{CS} C. S. Jia, P Gao, and X. L. Peng, J. Phys. A : Math. Gen. {\bf 39}, 7737 (2006).
\bibitem{AA} A. Arda and R. Sever, Commun. Theor. Phys. {\bf 64}, 269 (2015).
\bibitem{BDF} B. D. Figueiredo, I.D. Soares, J. Tiomno, Class. Quantum Grav. {\bf 9}, 1593 (1992).
\bibitem{ND} N. Drukker, B. Fiol and J. Simon, JCAP (2004) {\bf 10} : 012.
\bibitem{ND2} N. Drukker, B. Fiol and J. Simon, Phys. Rev. Lett. {\bf 91}, 231601 (2003). 
\bibitem{SD} S. Das and J. Gegenberg, Gen. Relativ. Grav. {\bf 40}, 2115 (2008).
\bibitem{LL} L. D. Landau and E. M. Lifshitz, {\it Quantum mechanics : non-relativistic theory}, Pergamon (2013).
\bibitem{LCNS} L. C. N. Santos and C. C. Barros Jr., EPJC (2018) {\bf 78} : 13.
\bibitem{LCNS2} L. C. N. Santos and C. C. Barros Jr., EPJC (2017) {\bf 77} : 186.
\bibitem{EPJC} F. Ahmed, Eur. Phys. J. C (2018) {\bf 78} : 598.
\bibitem{EPJC2} F. Ahmed, Eur. Phys. J. C (2019) {\bf 79} : 104.
\bibitem{EPJC3} F. Ahmed, Eur. Phys. J. C (2019) {\bf 79} : 534.
\bibitem{JC} J. Carvalho, A. M. de M. Carvalho and C. Furtado, EPJC (2014) {\bf 74} : 2935.
\bibitem{JC2} J. Carvalho, A. M. de M. Carvalho, E. Cavalcante and C. Furtado, EPJC (2016) {\bf 76}: 365.
\bibitem{GQG} G. Q. Garcia, J. R. de S. Oliveira, K. Bakke, C. Furtado, EPJ Plus {\bf 132}, 123 (2017).
\bibitem{GQG2} G. Q. Garcia, J. R. de S. Oliveira, C. Furtado, Int. J. Mod. Phys. D {\bf 27}, 1850027 (2018).
\bibitem{SGF} S. G. Fernandes, G. de A. Marques, V. B. Bezerra, Class. Quantum Grav. {\bf 23}, 7063 (2006).
\bibitem{AH} A. Havare, T. Yetkin, Class. Quantum Grav. {\bf 19}, 1 (2002). 
\bibitem{LCNS3} L. C. N. Santos, C. C. Barros Jr., EPJ C (2016) {\bf 76} : 560. 
\bibitem{MH} M. Hosseinpour, H. Hassanabadi and M. de Montigny, EPJ C (2019) {\bf 79}: 311.
\bibitem{MM} Marc de Montigny, S. Zare, and H. Hassanabadi, Gen. Relativ. Grav. (2018) {\bf 50} : 47.
\bibitem{EPJC4} P. Sedaghatnia, H. Hassanabadi and F. Ahmed, EPJ C (2019) {\bf 79} : 541.
\bibitem{HF} H. Fakhri and N. Abbasi, Jour. Math. Phys. {\bf 42}, 2416 (2001).
\bibitem{ASDC} A. S. de Castro, Phys. Lett. A {\bf 305}, 100 (2002).
\bibitem{JRH} J. R. Hiller, Amer. Jour. Phys. {\bf 70}, 522 (2002).
\bibitem{MHP} M. H. Pacheco, R. V. Maluf, C. A. S. Almedia and R. R. Landim, EPL {\bf 108}, 10005 (2014).
\bibitem{AOP} F. Ahmed, Ann. Phys. (N. Y.) {\bf 401}, 193 (2019).
\bibitem{AOP2} F. Ahmed, Ann. Phys. (N. Y.) {\bf 404}, 1 (2019).
\bibitem{GERG} F. Ahmed, Gen. Relativ. Grav. (2019) {\bf 51} : 69.
\bibitem{MDO} M. D. de Oliveria, A. G. M. Schmidt, Ann. Phys. (N. Y.) {\bf 401}, 21 (2019).
\bibitem{JC3} J. Carvalho, C. Furtado and F. Moraes, Phys. Rev. {\bf A 84}, 032109 (2011).
\bibitem{Gur} M. G\"{u}rses, Class Quantum Grav. {\bf 11}, 2585 (1994).
\bibitem{RJG} R. J. Gleiser, M. G\"{u}rses, A Karasu and O. Sarioglu, Class. Quantum Grav. {\bf 23}, 2653 (2006).
\bibitem{MG}  M. G\"{u}rses, Gen. Relativ. Grav. {\bf 42}, 1413 (2010).
\bibitem{MG2} M. G\"{u}rses, A Karasu and O. Sarioglu, Class. Quantum Grav. {\bf 22}, 1527 (2006).
\bibitem{Hassan} M. de Montigny, S. Zare and H. Hassanabadi, Gen Relativ. Grav. (2018) {\bf 50} : 47.


\end{thebibliography}
\end{document}